\documentclass[epj,twocolumn]{webofc}
\usepackage[varg]{txfonts}   

\woctitle{DHF 2014}

\begin{document}

\title{Experimental study of $\eta$ meson photoproduction reaction at MAMI}

\author{V.~L.~Kashevarov\inst{1,2}\fnsep\thanks{\email{kashev@kph.uni-mainz.de}}
for A2 Collaboration at MAMI}

\institute{Institut f\"ur Kernphysik, Johannes Gutenberg-Universit\"at Mainz,
 D-55099 Mainz, Germany
\and Lebedev Physical Institute, 119991 Moscow, Russia}

\abstract{
New data for the differential cross sections, polarization observables $T$, $F$, and $E$
in the reaction of $\eta$ photoproduction on proton from the threshold up to a center-of-mass
energy of W=1.9 GeV are presented.
The data were obtained with the Crystal-Ball/TAPS detector
setup at the Glasgow tagged photon facility of the Mainz Microtron MAMI.
The polarization measurements were made using a frozen-spin butanol target
and circularly polarized photon beam.
The results are compared to existing experimental data and different PWA
predictions.
The data solve a long-standing problem related the angular dependence of older
$T$ data close to threshold. The unexpected relative phase motion between 
$s$- and $d$-wave amplitudes required by the old data is not confirmed.
At higher energies, all model predictions fail to reproduce the new polarization
data indicating a significant impact on our understanding of the underlying
dynamics of $\eta$ meson photoproduction.
Furthermore, we present a fit of the new data and existing data from GRAAL
for $\Sigma$ asymmetry based on an expansion in terms of associated Legendre
polynomials.
A Legendre decomposition shows the sensitivity to small partial-wave contributions. 
The sensitivity of the Legendre coefficients to the nucleon resonance parameters   
is shown using the $\eta$MAID isobar model.
}

\maketitle

\section{Introduction}

The most baryon spectroscopy data have been obtained using $\pi N$ scattering data.
Pion photoproduction on nucleons is some additional tool for the investigation of
the nucleon resonances, especially in case of small $\pi N$ partial width.
Compared to pion, $\eta$ photoproduction has some additional advantages.
First, the $\eta NN$ coupling is very small. For example, this value of 
$g^2_{\eta NN}/4\pi = 0.4 \pm 0.2$ was obtained in Ref. \cite{Tiator94} in an
analysis of the angular distributions of $\eta$ photoproduction, that is by $\sim 30$ times
smaller than for pions. Second, because of the isoscalar nature of the $\eta$ meson, 
only nucleon excitations with isospin $I=1/2$ contribute to the $\gamma N \to \eta N$ reactions.
Both these factors simplify the extraction of the nucleon resonance parameters.

\begin{figure}
\begin{center}
\resizebox{0.48\textwidth}{!}{%
\includegraphics{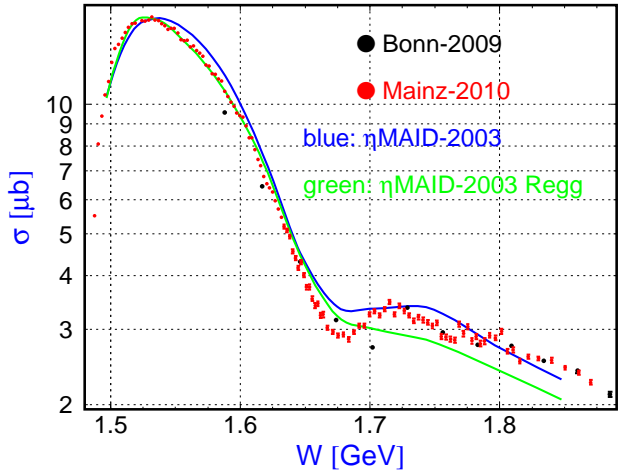}}
\caption{Total cross section of the $\gamma p \to \eta p$ reaction.
Black circles: Bonn data \cite{Crede09}, red circles: Mainz data \cite{McNicoll10}, 
blue curve: $\eta$MAID isobar model \cite{MAID}, 
green curve: reggeized $\eta$MAID isobar model \cite{MAIDr}.
}
\label{fig1}
\end{center}
\end{figure}

\begin{figure}
\begin{center}
\resizebox{0.48\textwidth}{!}{%
\includegraphics{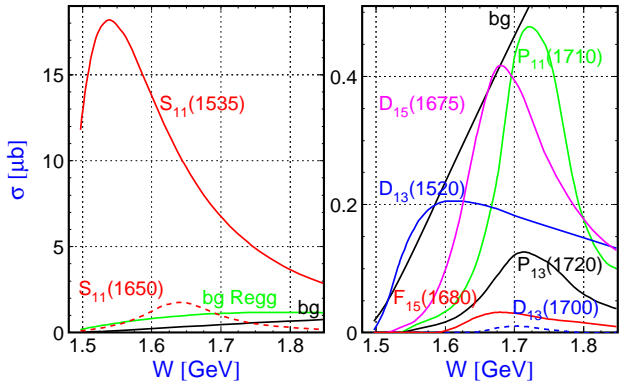}}
\caption{Partial contributions to the total cross sections from different resonances,
predicted by $\eta$MAID isobar model \cite{MAID} 
and non-resonance background for two $\eta$MAID versions: bg \cite{MAID} and 
bgRegg \cite{MAIDr}.
}
\label{fig2}
\end{center}
\end{figure}

The special feature of the $\gamma N \to \eta N$ reaction is the dominance of the $E_{0+}$ multipole
amplitude, which is populated by the $N^*(1535)1/2^-$ and $N^*(1650)1/2^-$ resonances. 
An interference between these resonances successfully explained a narrow structure in the
total cross section of $\eta$ photoproduction off the neutron \cite{BnGa14}. Experimental data
for the total cross section of the $\gamma p \to \eta p$ reaction together with two PWA predictions
are shown in Fig.\,\ref{fig1}. Partial resonance and non-resonance contributions to 
the total cross sections are shown in Fig.\,\ref{fig2} as an example of the $\eta$MAID predictions \cite{MAID},\cite{MAIDr}. 
The dominant role of the the $N^*(1535)1/2^-$ is illustrated in the left panel of the  
Fig.\,\ref{fig2}. Despite the fact that the Born terms give an insignificant contribution, a visible 
non-resonance background remains due to $\rho$ and $\omega$ exchange in the $t$-channel 
(black and green lines in the left panel for two version of the $\eta$MAID predictions).  
Other possible resonance contributions lie below the background (right panel).
Nevertheless these resonances can be identified by using the interference with the
dominant $E_{0+}$ multipole amplitude in the polarization observables.

In this paper, new experimental data for the $\gamma p \to \eta p$ reaction 
will be presented together with preliminary results of the partial-wave analysis based on
the Legendre fit to the data and the $\eta$MAID isobar model.       
\section{Experimental setup and data analysis}


\begin{figure}
\begin{center}
\resizebox{0.45\textwidth}{!}{%
\includegraphics{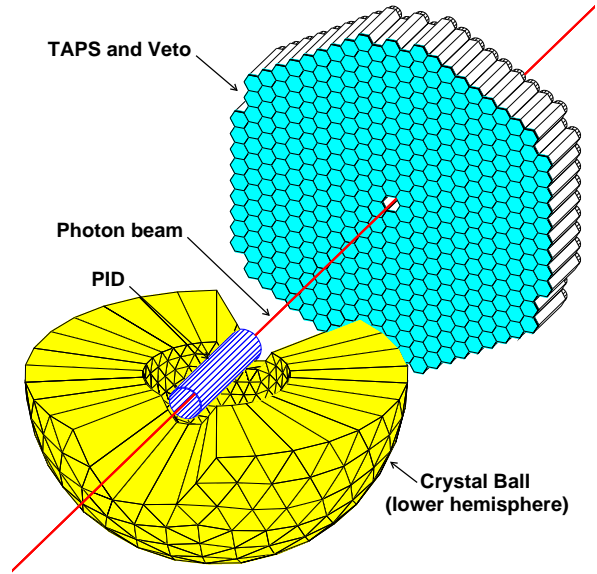}}
\caption{Experimental setup. The upper hemisphere of the Crystal Ball is omitted to show 
the inside region.}
\label{fig3}
\end{center}
\end{figure}

The experiment was performed at the MAMI C accelerator in Mainz\,\cite{MAMIC} using the
Glasgow-Mainz tagged photon facility\,\cite{TAGGER}. The quasi-monochromatic photon beam
covered the energy range from 700 to 1450 MeV.  
The experimental setup is shown schematically in Fig.\,\ref{fig3}. The bremsstrahlung 
photons, produced by the electrons in a $10\,\mu$m copper radiator and collimated by a
lead collimator, impinged on a target located in the center of the Crystal Ball detector\,\cite{CB}.
This detector consists of 672 optically isolated NaI(Tl) crystals with a thickness of 
15.7 radiation lengths covering 93\% of the full solid angle.
For charged-particle identification a barrel of 24 scintillation counters 
(Particle Identification Detector \,\cite{PID}) surrounding the target was used.
The forward angular range $\theta = 1 - 20^\circ$ is covered by the TAPS calorimeter
\,\cite{TAPS}. TAPS consists of 384 hexagonally shaped BaF2 detectors, each of which is
25 cm long, which corresponds to 12 radiation lengths. 
A 5-mm thick plastic scintillator in front of each module allows the identification 
of charged particles. 
The solid angle of the combined Crystal Ball and TAPS detection system is nearly 
$97\%$ of $4\pi$ sr.
More details on the energy and angular resolution of the CB and TAPS detector system are given
in Ref.\,\cite{Setup}.
 
In the polarization measurements, a longitudinally polarized electron beam with 
an energy of 1557 MeV and  a polarization degree of 80\% was used.
The longitudinal polarization of electrons is transferred to circular
polarization of the photons during the bremsstrahlung process in a radiator.
The degree of circular polarization depends on the photon energy and ranged from 65\%
at 700 MeV to 78\% at 1450 MeV \cite{Olsen}.

The experiment requires transversely (or longitudinally) polarized protons, which were provided by a
frozen-spin butanol ($\mathrm{C_4H_9OH}$) target.  
A specially designed $\mathrm{^3He/^4He}$ dilution refrigerator was built in order to maintain
a temperature of 25 mK during the measurements.
The target container, length 2 cm and diameter 2 cm, was filled with 2-mm diameter butanol
spheres with a packing fraction (filling factor) of around $60\%$.
The average proton polarization was $70\%$ with relaxation times of around 2000 h. 
The target polarization was measured at the beginning and the end of each data taking period. In order
to reduce the systematic errors, the direction of the target polarization vector was regularly
reversed during the experiment.
More details about the construction and operation of the target are given in Ref.\,\cite{Thomas}.

The mesons were identified via the $\eta \to 2 \gamma$ or $\eta \to 3 \pi^0 \to 6 \gamma$ decays.
Selections on the $2 \gamma$, or $6 \gamma$, invariant
mass distributions and on the missing mass $MM(\gamma p, \eta)$, calculated from the initial state and
the reconstructed $\eta$ meson, allowed for a clean identification of the reaction.
In order to subtract a background coming from quasi-free reactions on $\mathrm{^{12}C}$ and $\mathrm{^{16}O}$ nuclei 
of the butanol target, measurements on a pure carbon and a liquid hydrogen target were used.

\section{Results}

\begin{figure*}
\begin{center}
\resizebox{1.0\textwidth}{!}{%
\includegraphics{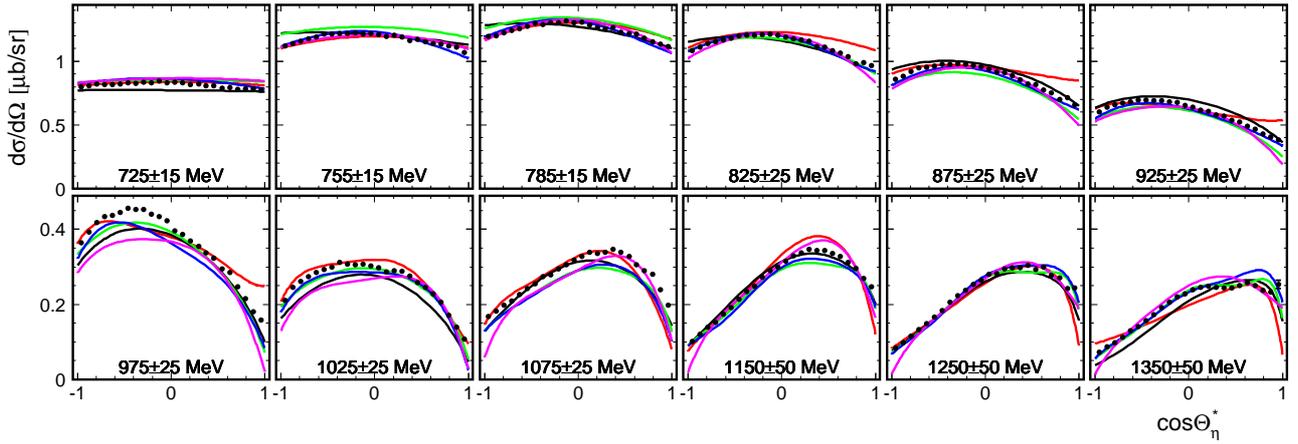}}
\caption{Differential cross sections. 
The new preliminary data with statistical uncertainties (black circles) are compared to
existing   
partial-wave analysis predictions (red lines: $\eta$-MAID \,\cite{MAID},
blue: SAID GE09 \,\cite{McNicoll10}).
green: BG2011-02 \,\cite{BnGa11},
black: Giessen model\,\cite{Giessen12},
magenta: Tryasuchev model\,\cite{Tryas}. 
The energy labels on the bottom of each panel indicate the photon energy bins
for our data.
}
\label{fig4}
\end{center}
\end{figure*}

Figure \,\ref{fig4} shows our preliminary results for differential cross sections 
together with various theoretical predictions   
\,\cite{MAID, McNicoll10, BnGa11, Giessen12, Tryas} for different bins in the incoming photon energy 
as a function of the $\eta$ meson polar angle in the center-of-mass system, $\theta_{\eta}^*$.
The present data agree well with previous measurements, but are much more precise. The original data have
a fine binning in energy, from 4 to 10 MeV, and span the full angular range. The data presented are averages 
over larger energy bins to be use for Legendre fits (see below). All model predictions are in reasonable 
agreement with the data. 

\begin{figure*}
\begin{center}
\resizebox{1.0\textwidth}{!}{%
\includegraphics{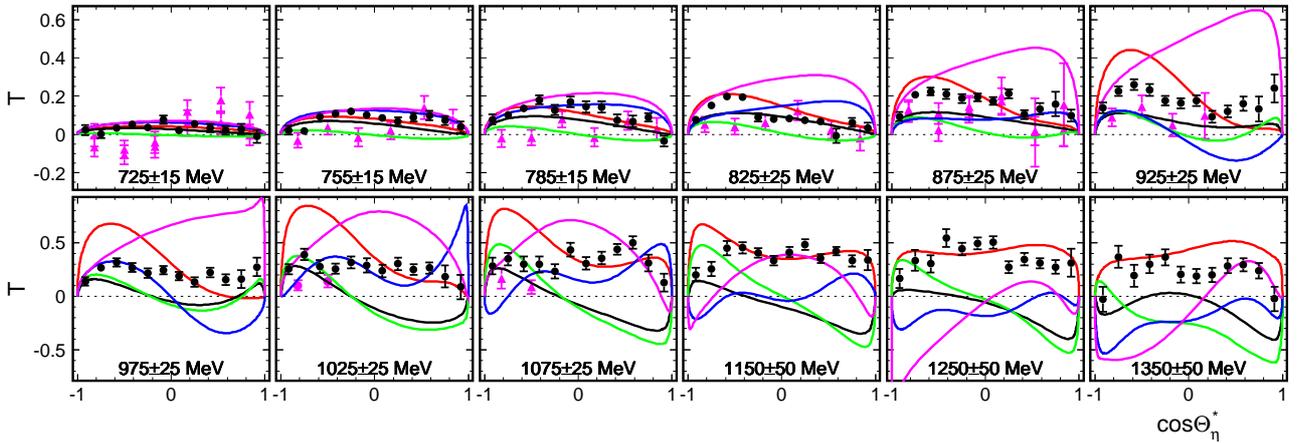}}
\caption{$T$ asymmetry.
The new Mainz results \cite{Mainz14} (black circles) are compared to
existing data from Bonn \cite{Bock1998} (magenta triangles).
Other notations same as in Fig.\,\ref{fig4}. 
}
\label{fig5}
\end{center}
\end{figure*}

\begin{figure*}
\begin{center}
\resizebox{1.0\textwidth}{!}{%
\includegraphics{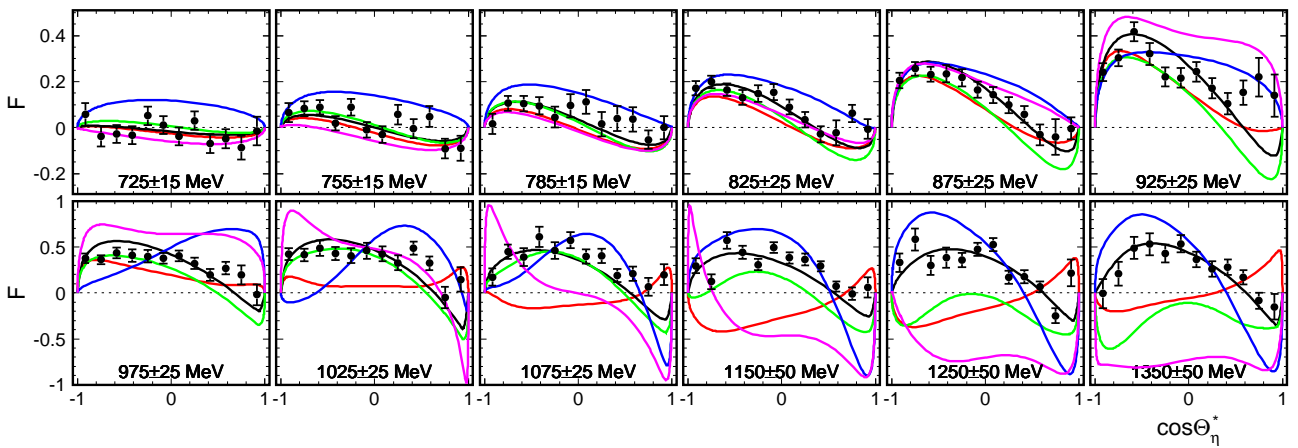}}
\caption{The new Mainz results \cite{Mainz14} for $F$ asymmetry.
Notations same as in Fig.\,\ref{fig4}.
}
\label{fig6}
\end{center}
\end{figure*}

Figures \,\ref{fig5} and \,\ref{fig6} show our results for $T$ and $F$ asymmetries \,\cite{Mainz14} 
together with previous data for $T$ \,\cite{Bock1998}.
The main inconsistencies with the existing data \cite{Bock1998} are in the near
threshold region. Here, our results do not confirm the observed nodal structure in the angular
dependence of the $T$ asymmetry and solve the long-standing question related to the relative phase
between $s$- and $d$-wave amplitudes. Our data do not require any additional phase shift beyond a
Breit-Wigner parametrization of resonances.
This important conclusion is corroborated by preliminary data from ELSA \cite{CBELSA}.      
At higher energies, all existing theoretical predictions of both $T$ and $F$
are in poor agreement among themselves and with our experimental data, even though they describe the
unpolarized differential cross sections well, see Fig. \,\ref{fig4} . 
The new data will therefore have a significant impact on the partial-wave structure of all models.

\section{Legendre analysis}

The full angular coverage of our new differential cross sections and polarization observables
allow us to perform a quality fit with the Legendre series truncated to a maximum orbital angular 
momentum $\ell_{\mathrm{max}}$:

\begin{eqnarray}
\frac{d\sigma}{d\Omega} &=&
\sum\limits_{n=0}^{2 \ell_{\mathrm{max}}} A^{\sigma}_{n}P^0_{n}(\cos\Theta_{\eta}),  \\\label{LegPol2}
T (F) \;\; \frac{d\sigma}{d\Omega} &=&
\sum\limits_{n=1}^{2 \ell_{\mathrm{max}}} A^{T(F)}_{n}P^1_{n}(\cos\Theta_{\eta}), \\\label{LegPol1}
E \;\; \frac{d\sigma}{d\Omega} &=&
\sum\limits_{n=0}^{2 \ell_{\mathrm{max}}} A^{E}_{n}P^0_{n}(\cos\Theta_{\eta}),  \\\label{LegPol3}
\Sigma \;\; \frac{d\sigma}{d\Omega} &=&
\sum\limits_{n=2}^{2 \ell_{\mathrm{max}}} A^{\Sigma}_{n}P^2_{n}(\cos\Theta_{\eta}),  \label{LegPol4}
\end{eqnarray}
where $P^m_{n}(\cos\Theta_{\eta})$ are associated Legendre polynomials.

\begin{figure*}
\begin{center}
\resizebox{1.0\textwidth}{!}{%
\includegraphics{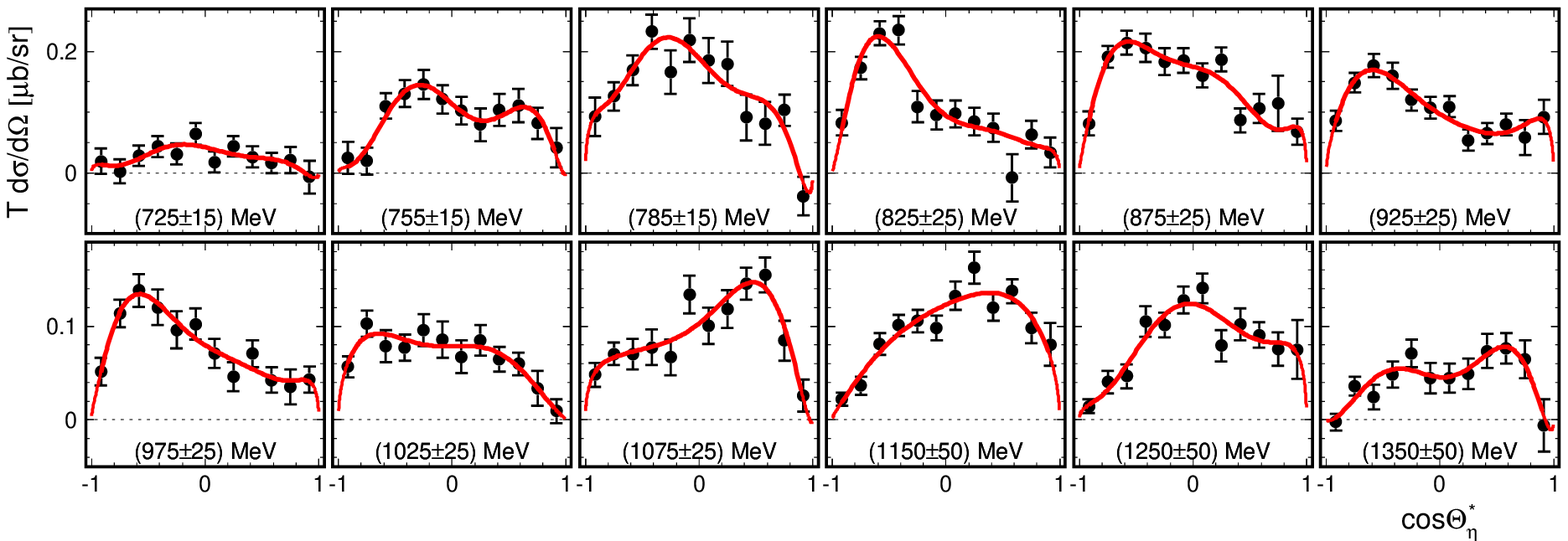}}
\caption{Our preliminary data for $T d\sigma/d\Omega$.
The result of the Legendre fit with $\ell_{\mathrm{max}} = 3$ is shown by the red curves.
}
\label{fig7}
\end{center}
\end{figure*}

\begin{figure*}
\begin{center}
\resizebox{1.0\textwidth}{!}{%
\includegraphics{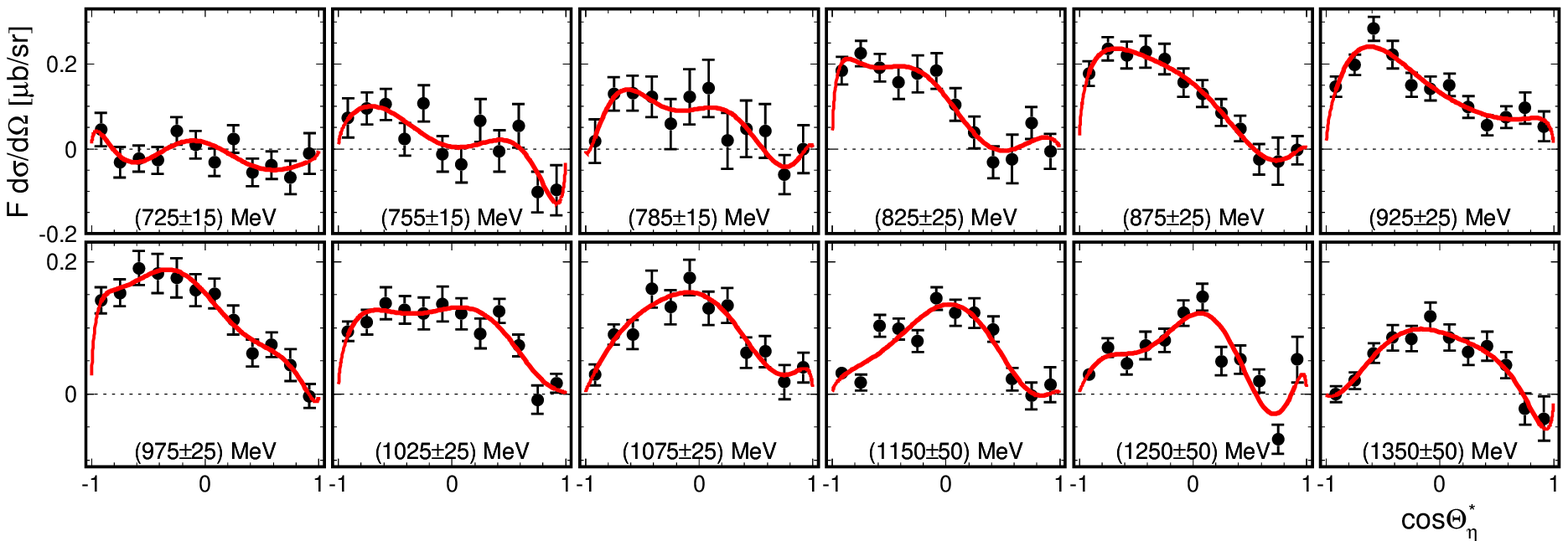}}
\caption{The same as Fig.\,\ref{fig7} for $F d\sigma/d\Omega$.
}
\label{fig8}
\end{center}
\end{figure*}

\begin{figure*}
\begin{center}
\resizebox{1.0\textwidth}{!}{%
\includegraphics{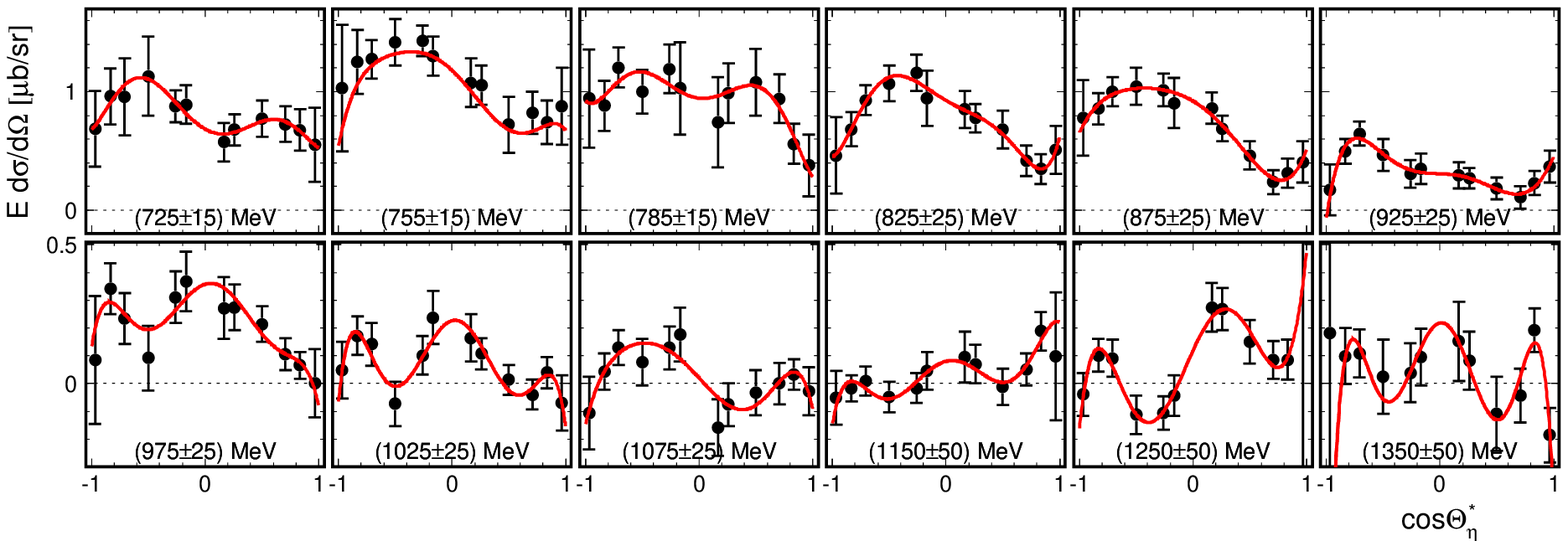}}
\caption{The same as Fig.\,\ref{fig7} for $F d\sigma/d\Omega$.
}
\label{fig9}
\end{center}
\end{figure*}

\begin{figure*}
\begin{center}
\resizebox{1.0\textwidth}{!}{%
\includegraphics{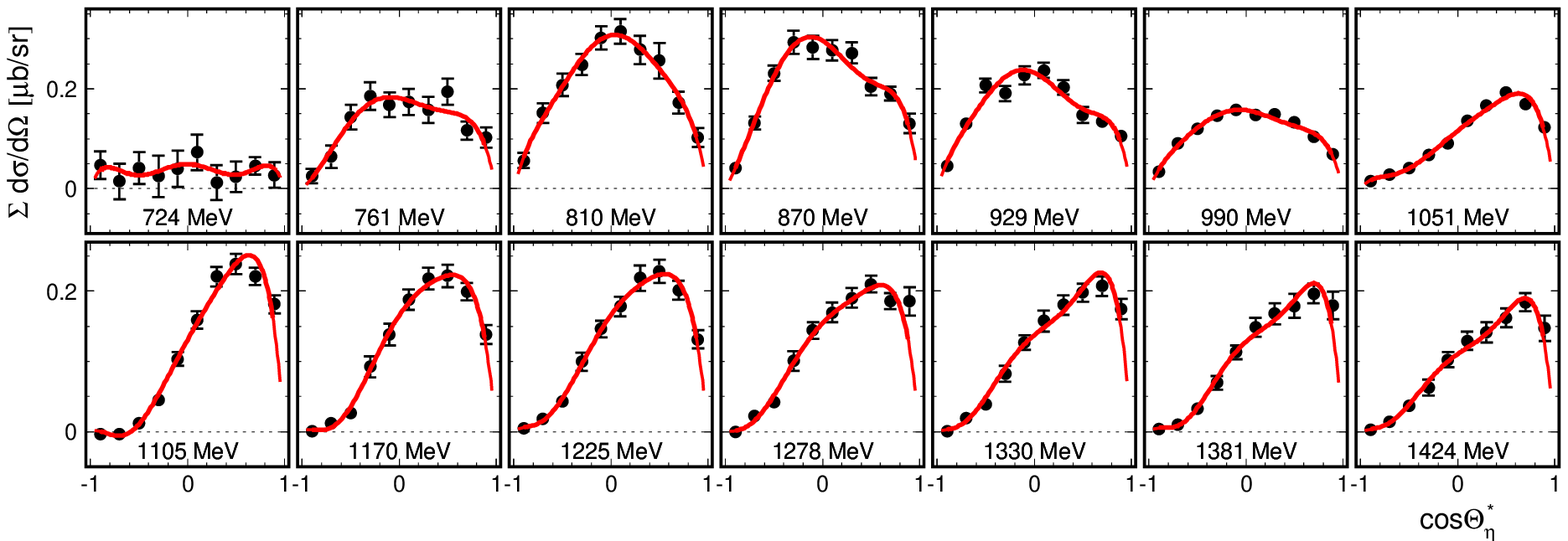}}
\caption{$\Sigma$ asymmetry \,\cite{Ajaka1998} multiplied by our new preliminary differential
cross sections. Red lines are the Legendre fit result with $\ell_{\mathrm{max}} = 3$. 
}
\label{fig10}
\end{center}
\end{figure*}

The spin-dependent cross sections, $T d\sigma/d\Omega$, $F d\sigma/d\Omega$, $E d\sigma/d\Omega$,
and $\Sigma d\sigma/d\Omega$ 
were obtained by multiplying the corresponding asymmetries with our new   
differential cross sections. Besides the observables $T$ and $F$, we used for the Legendre fit our 
preliminary data for the double polarization observable $E$ (circularly polarized photon beam and 
longitudinally polirized target) and the photon beam asymmetry $\Sigma$ (linerly polarized photon 
beam and unpolirized target) measured at the GRAAL facility \cite{Ajaka1998}.
Our preliminary data for the spin-dependent cross sections together with results of the Legendre fit 
with $\ell_{\mathrm{max}} = 3$ are shown in Figs \,\ref{fig7},\,\ref{fig8},\,\ref{fig9},\,\ref{fig10}.


\begin{figure*}
\begin{center}
\resizebox{1.0\textwidth}{!}{%
\includegraphics{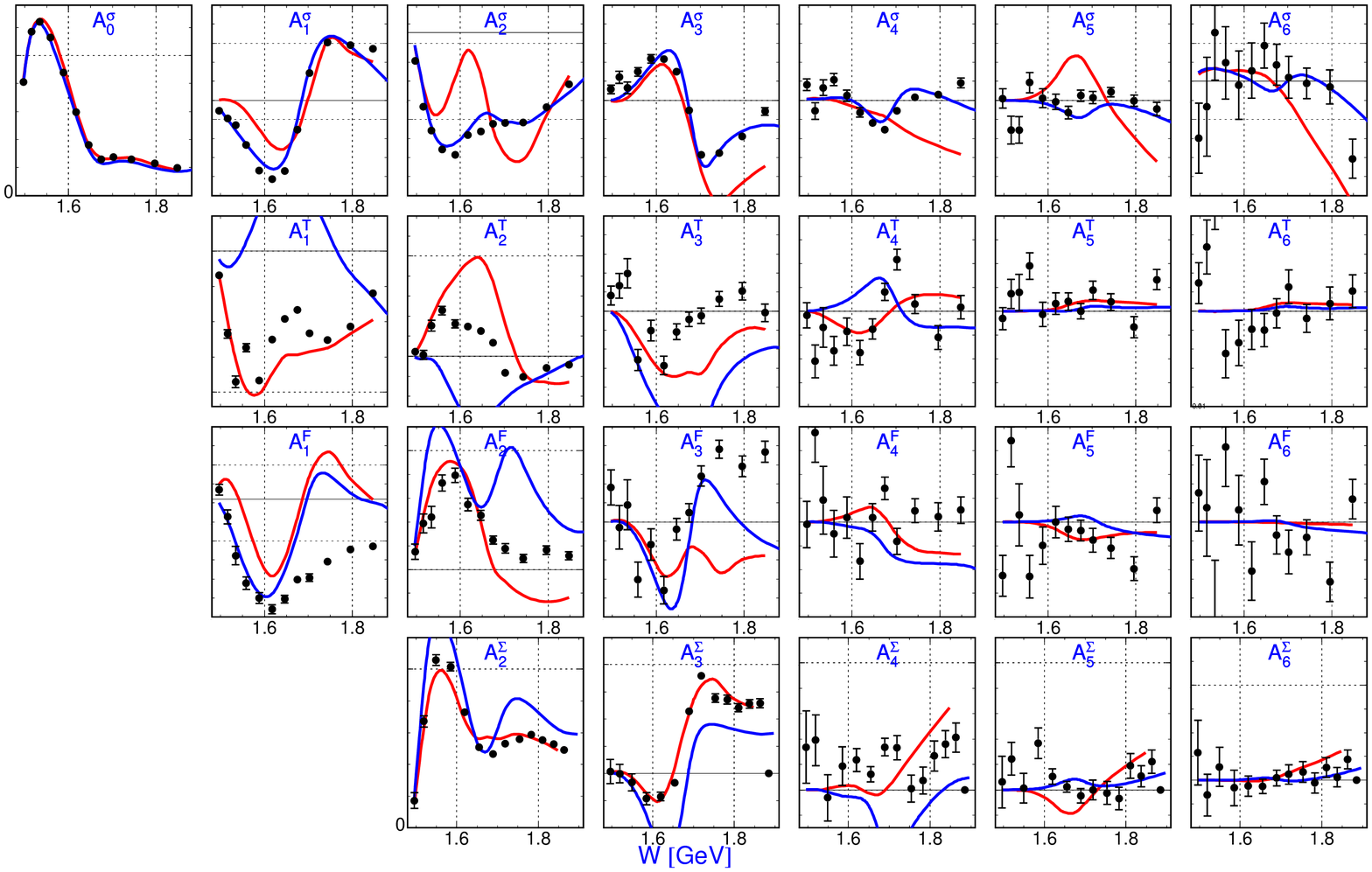}}
\caption{Legendre coefficients up to $\ell_{\mathrm{max}} = 3$ from
our fits to the different observables as function of the center-of-mass energy $W$ (black circles).
Red lines are the $\eta$MAID predictions. Blue lines are Solution 1. 
}
\label{fig11}
\end{center}
\end{figure*}

\begin{figure*}
\begin{center}
\resizebox{1.0\textwidth}{!}{%
\includegraphics{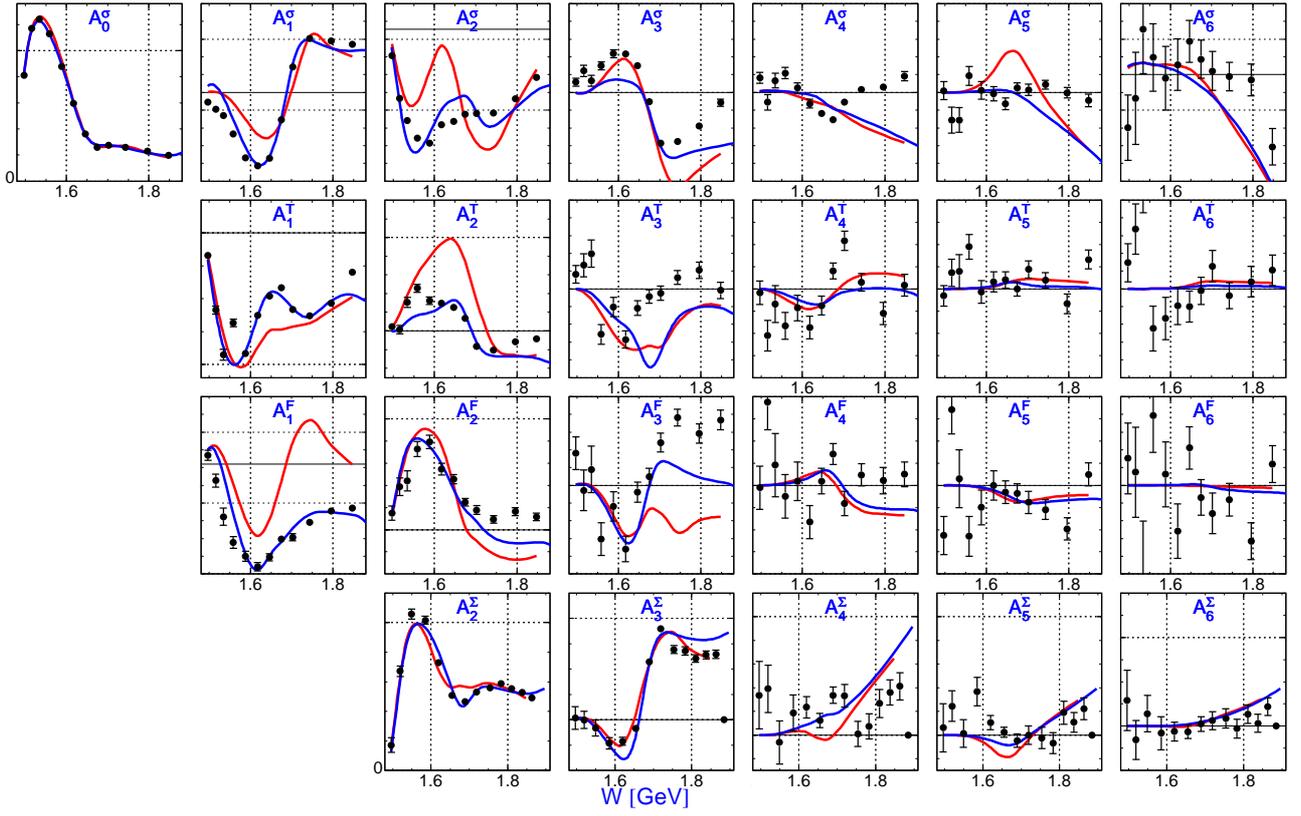}}
\caption{The same as Fig.\,\ref{fig11}, but blue lines are Solution 2.
}
\label{fig12}
\end{center}
\end{figure*}

The results for the Legendre coefficients
are presented in Figs.\,\ref{fig11} and \ref{fig12} (black circles).
\begin{figure*}
\begin{center}
\resizebox{1.0\textwidth}{!}{%
\includegraphics{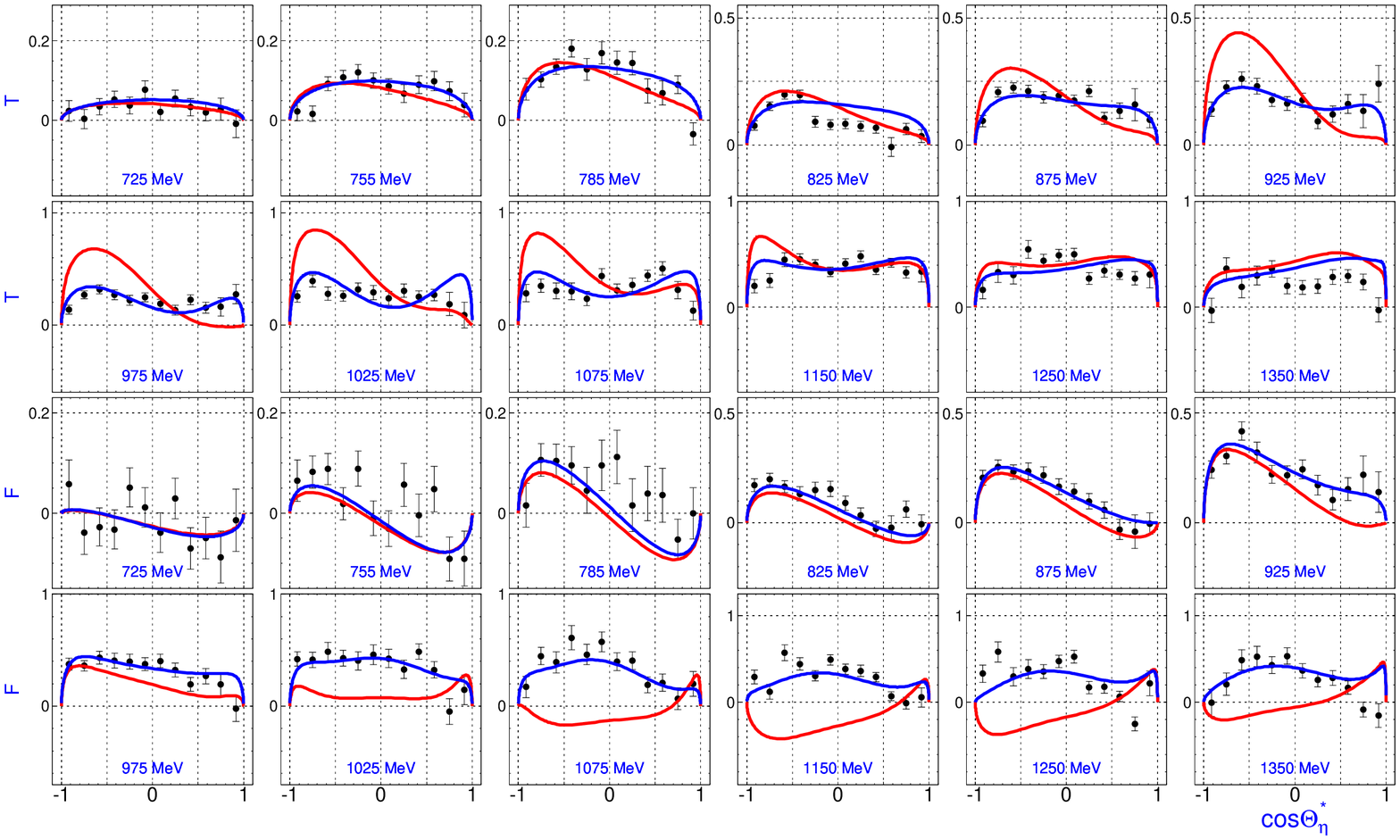}}
\caption{T and F asymmetries. Black circles: Mainz data \cite{Mainz14}.
Red lines: $\eta$MAID prediction \cite{MAID}. Blue lines: Solution 2. 
}
\label{fig13}
\end{center}
\end{figure*}
The last coefficient, $A_6$, depends only on $f$-wave contribution, $A_5$ is dominated by the
an interference between $d$ and $f$ waves, $A_4$ includes $d$, $f$ waves and an interference 
between $p$ and $f$ waves, and so on.
The first coefficient, $A^{\sigma}_0$, includes all possible partial-wave amplitudes and reflects
the magnitude of the total cross section. As expected the coefficient in a good agreement with 
the $\eta$MAID prediction (red line). The coefficients $A^{\Sigma}_n$ are also in reasonable agreement 
with the predictions, because the $\Sigma$ asymmetry was included to the $\eta$MAID fit. 

The fact that deviations of the $\eta$MAID prediction for the coefficients $A^{\sigma}_1$-$A^{\sigma}_5$ 
(top raw in Fig.\,\ref{fig11}) are mach larger than for the differential cross sections themselves (see Fig.\,\ref{fig4}) 
prompted us to involve these coefficients instead the differential cross sections in the data base for obtaining 
a new solution of the $\eta$MAID isobar model.
Results of the $\eta$MAID fit to the coefficients $A^{\sigma}_0$-$A^{\sigma}_6$ (Solution 1) are shown in Fig.\,\ref{fig11} 
as blue curves. Solution 1 significantly improved the description of the coefficients $A^{\sigma}_1$-$A^{\sigma}_6$,   
but ruined all others. Results of the $\eta$MAID fit to the all coefficients, 
$A^{\sigma}_n$, $A^{T}_n$, $A^{F}_n$, $A^{\Sigma}_n$, (Solution 2) are shown in Fig.\,\ref{fig12}.
This very preliminary solution is much better suited to describe the entire dataset, especially for the lowest 
coefficients, $A_1$, $A_2$.  
Probably involving additional resonances in the model will improve the situation with more high coefficients.
Here we just demonstrated the impact of the new data for future partial-wave analyses. 
New $\eta$MAID predictions based on Solution 2 for the observables $T$ and $F$ are shown in Fig.\,\ref{fig13} (blue lines).

\section{Summary}

In summary, we have presented new experimental data for the target asymmetry $T$, the
transverse beam-target observable $F$, preliminary data for the longitudinal beam-target observable $E$
and the differential cross sections for the $\gamma p \to \eta p$ reaction.
All existing solutions from various partial-wave analyses fail to reproduce the new polarization data.
A Legendre decomposition of the new results shows the sensitivity to small partial-wave contributions. 
We presented also results of the fit to the new data with the Legendre series truncated to a maximum 
orbital angular momentum $\ell_{\mathrm{max}}$.
Preliminary $\eta$MAID fit to the obtained Legendre coefficients results a new solution which much better
describes the new polarization data.  
Further improvement could be due to the addition of new resonances in the model, 
involving others polarization observables,
extending energy region for the data.

\section*{Acknowledgment}

This work was supported by the Deutsche Forschungsgemeinschaft (SFB 1044).

\newpage

\end{document}